\documentclass[aps,prl,reprint,superscriptaddress,nofootinbib]{revtex4-1}
\usepackage{graphicx}
\usepackage{hyperref}
\usepackage{amsmath,amssymb}

\begin{document}

\title{Consistency of Tachyacoustic Cosmology with de Sitter Swampland Conjectures}

\newcommand{\FIRSTAFF}{\affiliation{The Oskar Klein Centre for Cosmoparticle Physics,
	Department of Physics,
	Stockholm University,
	AlbaNova,
	10691 Stockholm,
	Sweden}}
\newcommand{\SECONDAFF}{\affiliation{Department of Physics,
	University at Buffalo, SUNY
	Buffalo,
	NY 14260
	USA}}

\author{Wei-Chen Lin}
\email{weichenl@buffalo.edu} 
\SECONDAFF
\author{William H. Kinney}
\email{whkinney@buffalo.edu}
\SECONDAFF
\FIRSTAFF

\date{\today}

\begin{abstract}
Recent studies show that there is tension between the de Sitter swampland conjectures proposed by Obeid, {\it et al.} and inflationary cosmology. In this paper, we consider an alternative to inflation, ``tachyacoustic'' cosmology, in light of swampland conjectures. In tachyacoustic models, primordial perturbations are generated by a period of superluminal sound speed instead of accelerating expansion. We show that realizations of tachyacoustic Lagrangians can be consistent with the de Sitter swampland conjectures, and therefore can in principle be consistent with a UV-complete theory. We derive a general condition for models with $c_S > 1$ to be consistent with swampland conjectures. 
\end{abstract}


\maketitle

\section{I. Introduction \label{sec1}}

String theory is so far the most prominent candidate 
of an ultraviolet (UV) complete theory of fundamental interactions, including gravity, which can be used to describe the early universe. However, it has recently been conjectured that the vast ``landscape'' of string vacua is surrounded by an even larger ``swampland'' of low-energy effective field theories (EFTs) which have no consistent UV completion \cite{Vafa:2005ui}. A number of conjectures have been proposed describing properties of low-energy EFTs which can be used to distinguish those with a consistent UV completion from ``swampland'' theories with no such completion, with perhaps the best studied being the Weak Gravity Conjecture \cite{ArkaniHamed:2006dz}. More recently, Refs. \cite{Obied:2018sgi,Agrawal:2018own} have conjectured that theories containing de Sitter vacua are in the swampland, and propose ``de Sitter swampland'' conjectures which must be satisfied for any consistent EFT limit of quantum gravity. The first conjecture is that there is an upper bound on the range transversed by scalar fields in field space 
\begin{equation}
\label{1.1}
\frac{|\Delta\phi|}{M_P}\lesssim\Delta\sim\mathcal{O}(1),
\end{equation}
where $ M_P=(8\pi G)^{-1/2} $ is the reduced Planck mass.  
The second conjecture \cite{Obied:2018sgi},  states that for UV-complete EFTs, there is a lower bound on the logarithmic gradient of the scalar field potential $ V(\phi) $
\begin{equation}
\label{1.2}
M_P\frac{|V'(\phi)|}{V}\gtrsim c \sim\mathcal{O}(1).
\end{equation}
The conjectures (\ref{1.1},\ref{1.2}) have been extensively studied in the context of inflationary cosmology  \cite{Agrawal:2018own,Kehagias:2018uem,Matsui:2018bsy,Achucarro:2018vey,Kinney:2018nny,Das:2018hqy,Ashoorioon:2018sqb,Brahma:2018hrd,Lin:2018rnx,Yi:2018dhl,Motaharfar:2018zyb,Dimopoulos:2018upl,Chiang:2018lqx,Lin:2018kjm,Holman:2018inr,Yi:2018dhl,Kinney:2018kew}. The general conclusion is that simple single-field inflation models are in strong tension with the swampland conjectures, while models beyond a single scalar field can in principle be consistent, for example Warm Inflation \cite{Berera:1995ie,Das:2018hqy,Das:2018rpg,Motaharfar:2018zyb}, or inflation involving generation of perturbations by an additional field \cite{Achucarro:2018vey}. Here, we offer a different point of view by asking if there are viable theories of the early Universe other than inflationary cosmology which are consistent with the swampland conjectures (\ref{1.1}) and (\ref{1.2}). Refs. \cite{Geshnizjani:2011dk,Geshnizjani:2011rm,Geshnizjani:2013lza,Geshnizjani:2014bya} considered general conditions on cosmology for the generation of primordial perturbations consistent with observation, and found four general classes of models, assuming standard General Relativity:
\begin{enumerate}
\item{A period of accelerated expansion ({\it i.e.} inflation).}
\item{A speed of sound faster than the speed of light.}
\item{Violation of the Null Energy Condition.}
\item{Inherently quantum-gravitational physics.}
\end{enumerate} 

In this paper, we consider models of the second class, in particular ``tachyacoustic'' cosmology, a type of k-essence model proposed by Magueijo \cite{Magueijo:2008pm} and by Bessada et al. \cite{Bessada:2009ns}.  As an alternative to inflation, tachyacoustic cosmology solves the horizon problem and creates a nearly scale invariant power spectrum via a superluminal speed of sound, $ c_S >1 $. (Although this special type of model involves superluminal sound speed, it has been argued that such fields do not violate causality and are as consistent as models with subluminal speed of sound \cite{Babichev:2007dw}.) We show that some (but not all) tachyacoustic cosmological models are consistent with the de Sitter swampland conjectures. In particular, we explain how the speed of sound plays the key role to determine whether or not a given model is consistent with swampland conjectures. Furthermore, we show that if the potential of the scalar field satisfies a certain general form, the two swampland criteria set the same condition on the tachyacoustic cosmology, {\it i.e.} consistency with one of the swampland conjectures guarantees consistency with the other.    

The rest of this paper is organized as follows. In Sec. II, we briefly review tachyacoustic cosmology. Section III is the main body of this paper, in which we derive general constraints on tachyacoustic models, and compare with the de Sitter swampland conjectures. In Sec. IV, we use the exact solution of two models introduced in Ref. \cite{Bessada:2009ns} to show that these general arguments hold in specific realizations of tachyacoustic Lagrangians. The conclusion is in Sec. V.  

\section{II. tachyacoustic cosmology \label{sec:review}}

In this section, we review the exact tachyacoustic solutions derived in Ref. \cite{Bessada:2009ns}, based on the generalization of the inflationary flow formalism \cite{Kinney:2002qn} introduced by  Bean et al. \cite{Bean:2008ga}. The attractor behavior and non-Gaussianity of these solutions were studied in Refs. \cite{Bessada:2012zx} and \cite{Bessada:2012rf}, respectively.

Consider a general Lagrangian of a scalar field with the form $ \mathcal{L}=\mathcal{L}[X, \phi] $, where $ X=\frac{1}{2}g^{\mu \nu}\partial_{\mu}\phi\partial_{\nu}\phi $ is the canonical kinetic term. Homogeneous modes of this scalar field form a perfect fluid with energy-momentum density
\begin{equation}
\label{2.1}
T_{\mu \nu}=(p+\rho)u_{\mu}u_{\nu}-pg_{\mu\nu},
\end{equation}
where
\begin{align}
\label{2.2}
p(X,\phi)&= \mathcal{L}(X,\phi),
\\
\label{2.3}
\rho(X,\phi)&= 2X\mathcal{L}_X-\mathcal{L}(X,\phi),
\\
\label{2.4}
u_{\mu}&= \frac{\partial_{\mu}\phi}{\sqrt{2X}}.
\end{align}
Here, the subscript "X" indicates a derivative with respect to the canonical kinetic term, $ \mathcal{L}_X \equiv \partial  \mathcal{L}/\partial X $. The sound speed is given by 
\begin{equation}
\label{2.5}
c^{2}_{S} \equiv \frac{p_X}{\rho_X}=\Bigg(1+2X\frac{\mathcal{L}_{XX}}{\mathcal{L}_X}\Bigg)^{-1},
\end{equation}
so that the canonical limit $ \mathcal{L}_{XX}=0 $ leads to $ c_{S}=1 $. 
We take a metric of the flat Friedmann-Robertson-Walker (FRW) form,
\begin{equation}
ds^2 = dt^2 - a^2\left(t\right) d {\bf x}^2.
\end{equation}
The Friedmann, Raychaudhuri, and continuity equation are
\begin{align}
\label{2.6}
H^2&=\frac{1}{3M_P^2}\rho=\frac{1}{3M_P^2}(2X\mathcal{L}_X-\mathcal{L}),
\\
\label{2.7}
\frac{\ddot{a}}{a}&=-\frac{1}{6M_P^2}(\rho+3p)= -\frac{1}{3M_P^2}(X\mathcal{L}_X +\mathcal{L}),
\\
\label{2.8}
\dot{\rho}&=-3H(\rho + p)=-6HX\mathcal{L}_X,
\end{align}
where $ a(t) $ is the scale factor and $ H=\dot{a}/a $ is the Hubble parameter, and the resulting equation of motion for the field $ \phi $ is 
\begin{equation}
\label{2.9}
\ddot{\phi}+3H\dot{\phi}+\frac{\dot{\phi}\partial_{0}\mathcal{L}_X}{\mathcal{L}_X}-\frac{\mathcal{L}_{\phi}}{\mathcal{L}_{X}}=0.
\end{equation}
For monotonic field evolution, one can instead use \textit{Hamilton Jacobi} approach \cite{Muslimov:1990be,Salopek:1990jq,Lidsey:1995np,Copeland:1993jj,Bean:2008ga}, in which the Hubble parameter $ H $ is treated as a fundamental quantity and all quantities are expressed as function of $ \phi $. In the homogeneous limit, $  \dot{\phi}=\sqrt{2X}$, the same dynamics is given by the master equation 
\begin{equation}
\label{2.10}
\dot{\phi}=\sqrt{2X}=-\frac{2M_{P}^2}{\mathcal{L}_{X}}H'(\phi),
\end{equation}
and the Hamilton Jacobi equation 
\begin{equation}
\label{2.11}
3M_{P}^2 H^2 (\phi)=\frac{4M_{P}^4 H'(\phi)^2}{\mathcal{L}_{X}}-\mathcal{L},
\end{equation}
where the prime denotes a derivative with respect to the field.  With the convention of the number of e-folds $ N $ defined as 
\begin{equation}
\label{2.12}
a(t)\propto exp \Big[\int_{t_{0}}^{t}Hdt\Big] \equiv e^{-N}, 
\end{equation}  
the differential $ dN $ can be re-written in terms of $ d\phi $ by:
\begin{equation}
\label{2.13}
dN=-Hdt=-\frac{H}{\sqrt{2X}}d\phi=\frac{\mathcal{L}_{X}}{2M_{P}^2} \bigg(\frac{H(\phi)}{H'(\phi)}\bigg)d\phi. 
\end{equation}  
Unlike the canonical case \cite{Kinney:2002qn}, which only contains a single hierarchy of Hubble slow roll parameters and the corresponding flow equations, in the k-essence generalization, one has to introduce three hierarchies of flow parameters and flow equations, since not only the Hubble parameter $ H(\phi) $ both the sound speed $ c_S $ and $ \mathcal{L}_{X}  $ can vary with time (for details, see Refs. \cite{Bean:2008ga,Bessada:2009ns}). The first parameters of these three sets of flow parameters are given by
\begin{align}
\label{2.14}
\epsilon(\phi)\equiv \frac{1}{H}\frac{dH}{dN}&=\frac{2M_{P}^2}{\mathcal{L}_{X}(\phi)}\bigg(\frac{H'(\phi)}{H(\phi)}\bigg)^2,
\\
\label{2.15}
s(\phi)\equiv -\frac{1}{c_S}\frac{dc_S}{dN}&=-\frac{2M_{P}^2}{\mathcal{L}_{X}(\phi)}\frac{H'(\phi)}{H(\phi)}\frac{c'_S(\phi)}{c_S(\phi)},
\\
\label{2.15.1}
\tilde{s}(\phi)\equiv \frac{1}{\mathcal{L}_{X}}\frac{d\mathcal{L}_{X}}{dN}&=\frac{2M_{P}^2}{\mathcal{L}_{X}(\phi)}\frac{H'(\phi)}{H(\phi)}\frac{\mathcal{L}_{X}'(\phi)}{\mathcal{L}_{X}(\phi)}.
\end{align}
Refs. \cite{Kinney:2007ag,Bessada:2009ns} derive a class of exact solutions to Eqs. (\ref{2.14},\ref{2.15},\ref{2.15.1}) such that the flow parameters $\epsilon$, $s$, and $\tilde s$ are all held constant, and the hierarchies of flow equations are satisfied exactly to all order.
In this class, the quantities $ H, c_S $, and $ \mathcal{L}_{X} $ can be solved exactly as 
\begin{equation}
\begin{aligned}
\label{2.16}
H & = H_0 e^{\epsilon N},
\\
c_S & = e^{-s N},
\\
\mathcal{L}_{X} &=  Ae^{\tilde{s} N},
\end{aligned}
\end{equation}
in which the number of e-folds is chosen to be zero when the sound speed is equal to one, and $ H_0 \equiv H(N=0) $ and $ A \equiv \mathcal{L}_{X}(N=0)  $ are the corresponding values of the Hubble parameter and $ \mathcal{L}_{X} $ respectively. Substituting $ H(N) $ and $ \mathcal{L}_{X}(N) $ from Eq. (\ref{2.16}) into Eq. (\ref{2.13}), one can solve for the evolution of field as a function of the number of e-folds  
\begin{equation}
\label{2.21}
\frac{\phi}{\phi_0}=e^{-\tilde{s}N/2},
\end{equation}
where $\phi_0$ is the field value when $c_S = 1$, and the coefficient $ A $ and $ \phi_0 $ are related as
\begin{equation}
\label{2.21b}
A=\frac{8 M_P^2 \epsilon}{\tilde{s}^2\phi_0^2}.
\end{equation}
From Eqs. (\ref{2.16}, \ref{2.21}, \ref{2.21b}), the quantities $ H$, $ c_S $ and $\mathcal{L}_{X}$ in terms of $  \phi$ are
\begin{align}
\label{2.17}
H & = H_0\bigg(\frac{\phi}{\phi_0}\bigg)^{-2\epsilon/\tilde{s}},
\\
\label{2.18}
c_S & = \bigg(\frac{\phi}{\phi_0}\bigg)^{2s/\tilde{s}},
\\
\label{2.19}
\mathcal{L}_{X} & = \frac{8\epsilon}{\tilde{s}^2}\bigg(\frac{M_P}{\phi}\bigg)^2.
\end{align}
From Eqs. (\ref{2.16}, \ref{2.12}), one can show that the background evolution is a power law 
\begin{equation}
\label{2.20}
a \propto e^{-N} \propto t^{1/\epsilon},
\end{equation}  
so the comoving Hubble horizon is proportional to the conformal time 
\begin{equation}
\label{2.22}
d_H \propto (aH)^{-1} \propto e^{(1-\epsilon)N} \propto \tau,
\end{equation}
and the acoustic horizon evolves as
\begin{equation}
\label{2.23}
D_H \propto \frac{c_S}{aH} \propto e^{(1-\epsilon-s)N} \propto \tau^{(1-\epsilon-s)/(1-\epsilon)}. 
\end{equation}
Note that the condition of a shrinking acoustic horizon is $ 1-\epsilon -s >0 $, which is different from a shrinking Hubble horizon $ \epsilon <1 $. The curvature perturbation is generated at the acoustic horizon, which is larger than the Hubble horizon when $ c_S>1 $, and therefore the horizon problem can be solved in a non-inflationary expansion scenario. From Eqs. (\ref{2.5}, \ref{2.18},\ref{2.19}), one can write the speed of sound $ c_S $ in terms of $ \mathcal{L}_X $: 
\begin{equation}
\label{2.24}
c^{2}_{S}=\Bigg(1+2X\frac{\mathcal{L}_{XX}}{\mathcal{L}_X}\Bigg)^{-1}=C^{-1}\mathcal{L}_X^{-2s/\tilde{s}},
\end{equation}
where C is defined as
\begin{equation}
\label{2.25}
C \equiv \Bigg(\frac{\tilde{s}^2\phi_0^2}{8M_P^2\epsilon}\Bigg)^{2s/\tilde{s}}= A^{-2s/\tilde{s}}.
\end{equation}
Note that the constant $ C $ is not a free parameter, but is fixed to be unity by the condition that the Lagrangian reduces to the canonical form $ \mathcal{L}=X-V $ in the limit $ c_S \rightarrow 1 $, as is evident from Eq. (\ref{2.24}). This condition sets the energy scale of $ \phi_0 $ to be of order $ M_P $. The right-hand side of Eq. (\ref{2.24}) can be expressed as a differential equation of $ \mathcal{L}(X, \phi) $,
\begin{equation}
\label{2.26}
2X\mathcal{L}_{XX}+\mathcal{L}_{X}-C\mathcal{L}_{X}^n=0,
\end{equation}
where n is defined as
\begin{equation}
\label{2.27}
n \equiv 1+\frac{2s}{\tilde{s}}.
\end{equation}
Therefore, one can construct a Lagrangian based on a given relationship between the parameters $ s $ and $ \tilde{s} $. The tachyacoustic solutions investigated in \cite{Bessada:2009ns} are $ n=0 $ (a Cuscuton-like model), and $ n=3 $ (a Dirac-Born-Infeld (DBI)-like model). Below, we summarize the results derived in Ref. \cite{Bessada:2009ns}: the reader is referred to this paper for details. 

The case $ n=0 $ or equivalently $ \tilde{s}=-2s $, has a Lagrangian of the form 
\begin{equation}
\label{2.28}
\mathcal{L}(X, \phi)=2f(\phi)\sqrt{X}+CX-V(\phi),
\end{equation}
where
\begin{align}
\label{2.29}
f(\phi)&=\frac{\sqrt{2}M_P^2H(\phi)\epsilon}{s\phi_0  c_S(\phi)}[c_S^2(\phi)-1],
\\
\label{2.30}
V(\phi)&=M_P^2H^2(\phi)\Big[3-\frac{\epsilon}{c_S^2(\phi)}\Big].
\end{align}
This has the form of a ``Cuscuton'' theory \cite{Afshordi:2006ad,Afshordi:2007yx,Boruah:2017tvg,Boruah:2018pvq}, with a kinetic term linear in the field velocity. We keep the constant $ C $ as an parameter in Eq. (\ref{2.28}) and implicitly in $ f(\phi) $ (\ref{2.29}) since this is the form used in Ref. \cite{Bessada:2009ns}. In the later discussion we will set $ C=1 $. The Hubble parameter and sound speed are given by:
\begin{align}
\label{2.31}
H(\phi)&=H_0\Big(\frac{\phi}{\phi_0}\Big)^{\epsilon/s},
\\
\label{2.32}
c_S(\phi)&=\Big(\frac{\phi_0}{\phi}\Big).
\end{align}
The Lagrangian of the DBI-like model, with $ \tilde{s}=s $, or $ n=3 $ and $ C=1 $ is
\begin{equation}
\label{2.29c}
\mathcal{L}(X, \phi)=-f^{-1}(\phi)\sqrt{1-2f(\phi)X}+f^{-1}(\phi)-V(\phi),
\end{equation}
with
\begin{align}
\label{2.33}
f(\phi)&=\Big(\frac{1}{2M_P^2\epsilon}\Big)\frac{1-c_S^2(\phi)}{H^2(\phi)c_S(\phi)},
\\
\label{2.34}
V(\phi)&=3M_P^2H^2(\phi)\Big[1-\Big(\frac{2\epsilon}{3}\Big)\frac{1}{1+c_S(\phi)}\Big].
\end{align}
The Hubble parameter and sound speed are given by:
\begin{align}
\label{2.31b}
H(\phi)&=H_0\Big(\frac{\phi}{\phi_0}\Big)^{-2\epsilon/s},
\\
\label{2.32b}
c_S(\phi)&=\Big(\frac{\phi}{\phi_0}\Big)^2.
\end{align}

Within the framework of constant flow parameters, the cosmological perturbation can be solved exactly at the linear level \cite{Bessada:2009ns}. The scalar spectral index of perturbations for a tachyacoustic solution is given by 
\begin{equation}
\label{3.01}
n_s=1-\frac{2\epsilon+s}{1-\epsilon-s}, 
\end{equation}
which has a scale invariant limit, $ s=-2\epsilon $. Radiation-dominated tachyacoustic expansion, which we will take here as a fiducial model, has $ \epsilon=2 $, with spectral index  
\begin{equation}
\label{3.02}
n_s=1+\frac{4+s}{1+s}. 
\end{equation}

\section{III. tachyacoustic cosmology tested by swampland criteria \label{sec:main}}

In this section, we derive general scaling rules relating tachyacoustic Lagrangians with the de Sitter swampland conjecture (\ref{1.1}). We then derive a general relation between models which satisfy the condition (\ref{1.1}) and the second de Sitter swampland conjecture (\ref{1.2}) for the case of potentials having the leading order behavior of a power law in the sound speed, which is the case considered here.  

To generate perturbations on a range of scales $k$ consistent with observation, we must have a sufficient number of e-folds of evolution, which from Eq. (\ref{2.23}) results in a bound $\Delta N \gtrsim \mathcal{O}(10)$. (Our results are not sensitive to the exact number of e-folds assumed.) We can obtain a general relation between $\Delta N$ and the field excursion $\Delta \phi$ using Eqs. (\ref{2.13},\ref{2.14}),
\begin{equation}
\label{3.a}
\Delta N \sim\pm \sqrt{\frac{\mathcal{L}_X}{2\epsilon}}\frac{\Delta \phi}{M_P}, 
\end{equation}
where the plus sign corresponds to $ \dot{\phi} <0 $ and the minus sign corresponds to $\dot{\phi} > 0$. For $ \phi $ positive, the DBI-like model has $ \dot{\phi}<0 $ and the Cuscuton-like model has $ \dot{\phi}>0 $.
Substituting Eq. (\ref{2.24}) with $ C=1 $ into Eq. (\ref{3.a}), we can make $ c_S $ explicit in the relation between the field excursion and the number of e-folds as
\begin{equation}
\label{3.aa}
\frac{\Delta \phi}{M_P} \sim \pm \sqrt{2\epsilon} (c_S)^{1/\beta} \Delta N, 
\end{equation}
where $ \beta \equiv 2s/\tilde{s} $, which for the Cuscuton-like model is
\begin{equation}
\label{3.b}
\frac{\Delta \phi}{M_P} \sim -\frac{\sqrt{2\epsilon}}{c_S}\Delta N, 
\end{equation}
and for the DBI-like model is
\begin{equation}
\label{3.c}
\frac{\Delta \phi}{M_P} \sim \sqrt{2\epsilon c_S}\Delta N. 
\end{equation}
We immediately see that the Cuscuton-like model satisfies the criterion (\ref{1.1}), since $\Delta\phi \propto c_S^{-1} \Delta N$ with $c_S \gg 1$. By contrast, the DBI-like model has $\Delta\phi \propto \sqrt{c_S} \Delta N$, which fails the criterion (\ref{1.1}) in the limit $c_S \gg 1$. \footnote{Note that here, unlike in the case of inflation, $\epsilon \sim \mathcal{O}\left(1\right)$.}  Moreover, Eq. (\ref{3.aa}) shows generally that the condition to pass the first swampland criterion (\ref{1.1}) is $ \beta=2s/\tilde{s} <0 $.

We can relate this in a general way to the second criterion \ref{1.2} as follows: If we consider the scale invariant limit in the case of a radiation-dominated background, {\it i.e.}  $ s=-4 $, from Eq. (\ref{2.16}) we can see that the sound speed drops rapidly as 
\begin{equation}
\label{3.14}
c_S=e^{4N},
\end{equation}
so that it is consistent to take $ c_S \gg \mathcal{O}(1) $. Then we can rewrite the logarithmic gradient of the scalar field potential in terms of the derivative with respect to sound speed as
\begin{equation}
\label{3.15}
M_P\frac{|V'(\phi)|}{V}=M_P\frac{1}{V}\left|\frac{dc_S}{d\phi}\frac{dV}{dc_S}\right|.
\end{equation}
Now, if the leading-order behavior of the potential is a power law in $c_S$ (which is the case for all potentials considered here), then the leading-order behavior of the derivative is
\begin{equation}
\label{3.15c}
 \frac{1}{V}\frac{dV}{dc_S} \sim \frac{1}{c_S}.
\end{equation}
From the relation (\ref{2.18}),
\begin{equation}
\label{3.16}
c_S = \bigg(\frac{\phi}{\phi_0}\bigg)^{2s/\tilde{s}} \sim \bigg(\frac{\phi}{M_P}\bigg)^{2s/\tilde{s}},
\end{equation}
so that we have 
\begin{equation}
\label{3.17}
M_P \frac{dc_S}{d\phi} \sim c_S^{(\beta-1)/\beta},
\end{equation}
where $ \beta = 2s/\tilde{s} $. Substituting Eq. (\ref{3.15c}) and Eq. (\ref{3.17}) into Eq. (\ref{3.15}) gives
\begin{equation}
\label{3.18}
M_P\frac{|V'(\phi)|}{V}\sim c_S^{-1/\beta},         
\end{equation}
which shows that the condition to pass the second swampland criterion (\ref{1.2}) is also to have $ \beta=2s/\tilde{s} <0 $. From Eq. (\ref{3.16}) we also have
\begin{equation}
\label{3.19}
\phi=\phi_0c_S^{1/\beta} \sim M_P c_S^{1/\beta}.
\end{equation}
Therefore, for any potential having the leading order behavior of a power law in $c_S$, the condition to satisfy {\it both} swampland conjectures is $ \beta=2s/\tilde{s} <0 $. The Cuscuton-like model, having $ \beta=-1 $, passes both of the swampland criteria but the DBI-like model, having $ \beta=2, $ does not. In the next section, we use the algebraic formulas in Sec. II to confirm this result.
 
\section{IV. testing exact solutions: the Cuscuton-like and the DBI-like models}

In this section, we examine in more detail swampland constraints on the models discussed in Sec. II. It is convenient to use the variable $ x $ defined as $ x \equiv \phi/M_P $, and a constant $ \alpha \equiv \phi_0 /M_P \sim \mathcal{O}\left(1\right)$ to express the field in units of the reduced Planck mass directly. In terms of variable $ x $, the first swampland conjecture Eq. (\ref{1.1}) reads as
\begin{equation}
\label{3.3a}
|\Delta x|\lesssim\Delta\sim\mathcal{O}(1), 
\end{equation} 
and the second swampland conjecture Eq. (\ref{1.2}) reads as  
\begin{equation}
\label{3.6}
\frac{|V'(x)|}{V}\gtrsim c \sim\mathcal{O}(1).
\end{equation}

\subsection{A. the Cuscuton-like model: $ \tilde{s}=-2s $}

Substituting the Cuscuton condition $ \tilde{s}=-2s $ and $ s = -4 $ into Eq. (\ref{2.21}), the field evolves as 
\begin{equation}
\label{3.1}
\phi=\phi_0 e^{-4N}=\alpha M_P e^{-4N}.
\end{equation}
In terms of the variable $ x\equiv \phi/M_P $ and  $ \alpha=1/2 $ from the condition of $ C=1 $ , Eq. (\ref{3.1}) becomes
\begin{equation}
\label{3.3}
x\equiv \frac{\phi}{M_P}= \frac{1}{2} e^{-4N}.
\end{equation}
Remembering that early time and $c_S \gg 1$ correspond to $N \rightarrow +\infty$, we consider exclusively the region $N > 0$. (In fact, from Eq. (\ref{2.30}) it's easy to show that the potential quickly becomes negative once the sound speed is less than the speed of light. To ensure constency with cosmology, the tachyacoustic field must decay into standard model particles before reaching this point. See Ref. \cite{Bessada:2009ns} for discussion.) From Eq. (\ref{3.3}) we see that the range traversed of the field from a given $ N $ to the point $ c_S=1 $, which we take to be the endpoint of tachyacoustic evolution, is 
\begin{equation}
\label{3.3b}
|\Delta x |= \frac{1}{2} (1-e^{-4N}),
\end{equation}
which shows that the Cuscuton-like model passes the first swampland criterion, Eq. (\ref{3.3a}), and confirms the previous estimate in Eq. (\ref{3.b}).  
   
\begin{figure}[h]
	\hspace{30px}
	\centering
	\includegraphics[scale=0.30]{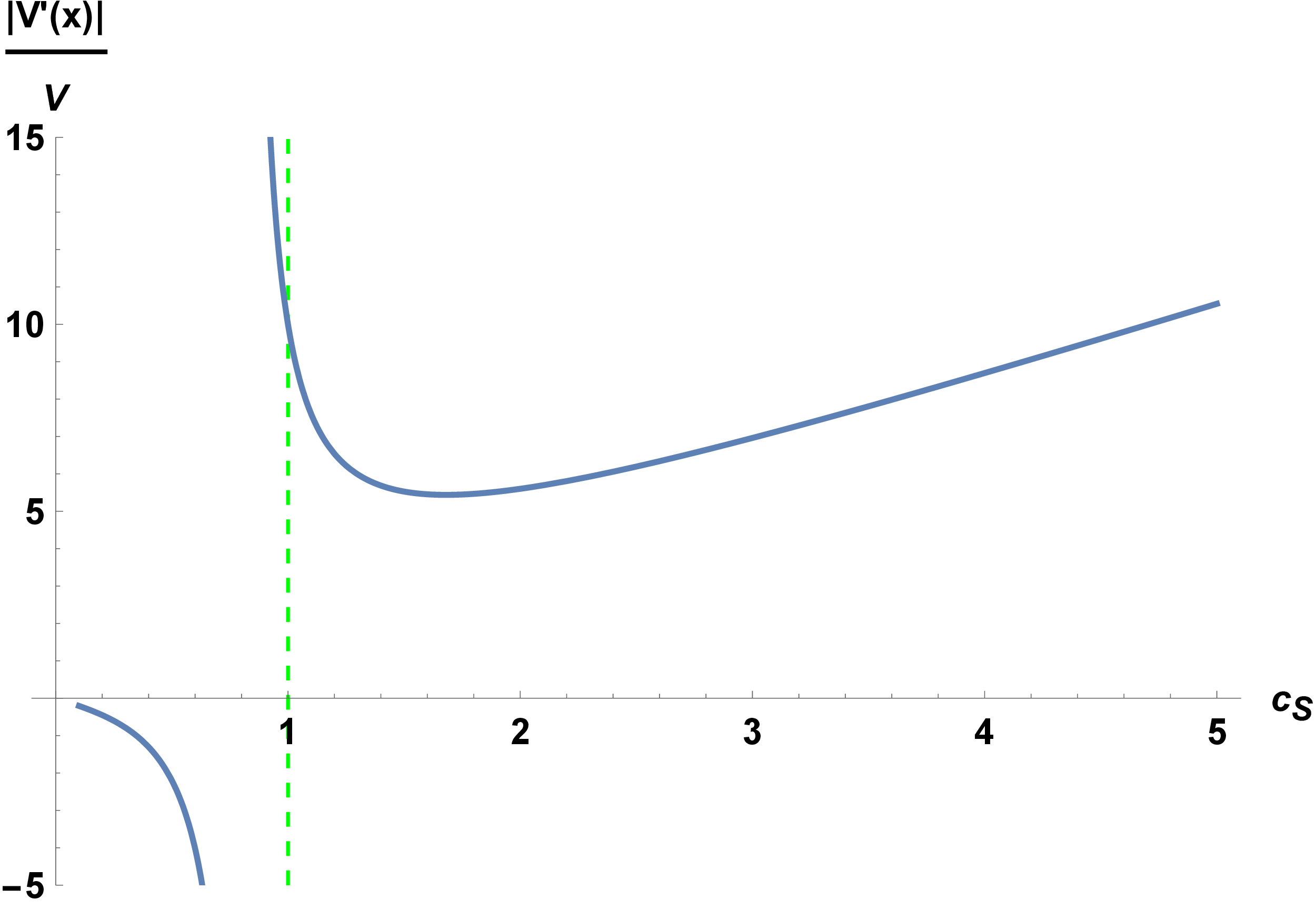}
	\caption{In the Cuscuton-like model, if the region traversed by the scalar field is restricted to $ c_S>1 $, the second swampland conjectures is satisfied, $ \frac{|V'(x)|}{V} \gtrsim \mathcal{O}(1) $. This figure also shows that the potential quickly becomes negative when $ c_S<1 $. The dashed (green) line indicates $ c_S=1 $. }
	\label{fig:1.0}
\end{figure}

To test the second swampland conjectures, we rewrite the potential of the Cuscuton-like model (\ref{2.30}) as
\begin{equation}
\label{3.4}
V(\phi)=3M_P^2H_0^2\Big(\frac{\phi}{\phi_0}\Big)^{2\epsilon/s}\Big[1-\frac{\epsilon}{3}\Big(\frac{\phi}{\phi_0}\Big)^2\Big], 
\end{equation}
and substitute $ \epsilon=2 $ and $ s=-4 $ to have
\begin{equation}
\begin{split}
\label{3.5}
V(\phi) &= 3M_P^2H_0^2\Big(\frac{\phi}{\phi_0}\Big)^{-1}\Big[1-\frac{2}{3}\Big(\frac{\phi}{\phi_0}\Big)^2\Big]  \\
&= 3M_P^2H_0^2 (2x)^{-1}\Big[1-\frac{2}{3}(2x)^2\Big].
\end{split}
\end{equation}
Eq. (\ref{3.5}) gives 
\begin{equation}
\label{3.7}
\frac{|V'(x)|}{V}  = \frac{ x|8+\frac{3}{x^2}|}{3-8x^2}=\frac{2c_S(3c_S^2+2)}{3c_S^2-2}, 
\end{equation} 
where Eq. (\ref{2.32}) is used in the second equality. A plot of $ |V'(x)|/V $ versus $ c_S $ is shown in Fig. \ref{fig:1.0} showing that the Cuscuton-like model is consistent with the criterion (\ref{1.2}) for $c_S \gg 1$.
In this limit, Eq. (\ref{3.7}) reduces to 
\begin{equation}
\label{3.7b}
\frac{|V'(x)|}{V} \sim 2 c_S= 2 e^{4N} \gg 1,
\end{equation} 
where, from Eq. (\ref{2.16}), we take $ s=-4 $. The Cuscuton-like model passes the second swampland criterion since $ |V'(x)|/V $ is roughly proportional to the sound speed, consistent with the general relation (\ref{3.18}).

\subsection{B. the DBI-like model: $ \tilde{s}=s $}
Again considering the scale-invariant limit $ \epsilon=2 $ and $s =-4$, but with  $ \tilde{s}=s $ for the DBI-like model, from Eq. (\ref{2.21}), the field evolves as 
\begin{equation}
\label{3.9}
x=\frac{\phi}{M_P} = \alpha e^{2N}.
\end{equation}
In the DBI case, the condition $ C=1 $ gives $ \alpha \equiv \phi_0/M_{\rm P} = 1 $. The field excursion from a given $ N $ to the point $ c_S=1 $ is 
\begin{equation}
\label{3.9b}
|\Delta x |=  (e^{2N}-1),
\end{equation}
which shows a strong violation of the first swampland criterion, Eq. (\ref{3.3a}), and confirms our estimate in Eq. (\ref{3.c}). 

\begin{figure}[h]
	\hspace{30px}
	\centering
	\includegraphics[scale=0.30]{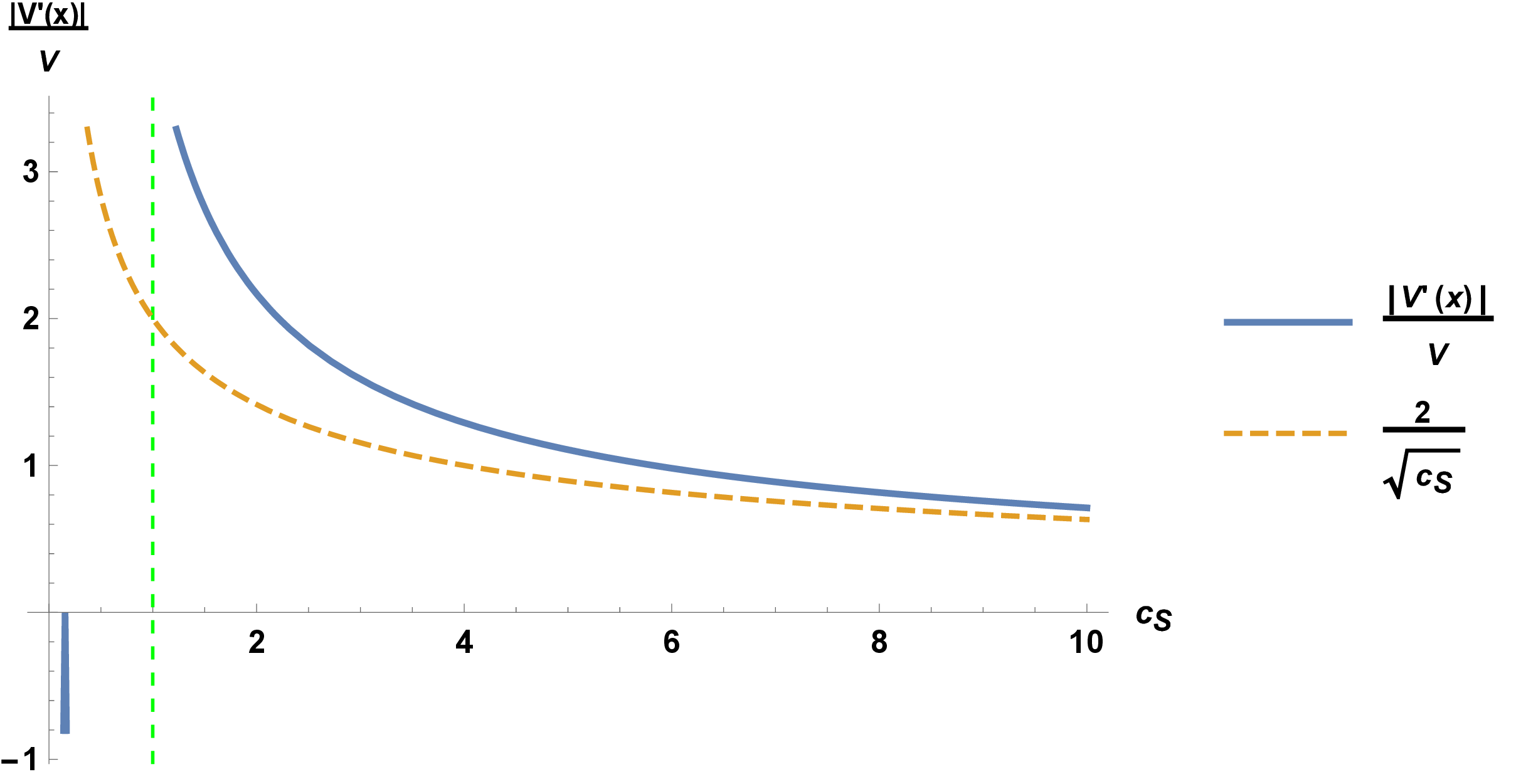}
	\caption{In the DBI-like model, the scalar field traverses a long distance in the region  $ c_S \gg1  $, where the second swampland criterion is strongly violated, $ \frac{|V'(x)|}{V} \ll \mathcal{O}(1) $. The solid (blue) line, $ \frac{|V'(x)|}{V} $, approaches the dashed (orange) line, which is the limiting case $ \frac{2}{\sqrt{c_S}} $, when $ c_S \gg1  $. }
	\label{fig:2.0}
\end{figure}

Next, to check the second swampland criterion we rewrite the potential (\ref{2.34}) as 
\begin{equation}
\label{3.10}
V(\phi)=3M_P^2H_0^2\Big(\frac{\phi}{\phi_0}\Big)^{-4\epsilon/s}\Big[1-\frac{2\epsilon}{3}\frac{1}{1+\big(\frac{\phi}{\phi_0}\big)^2}\Big], 
\end{equation}
then substitute $ \epsilon=2 $, $ s=-4 $ and $ \phi_0 = M_P $ together with the change of variable $ x \equiv \phi/M_P $ to get
\begin{equation}
\label{3.11}
V(x) = 3M_P^2H_0^2 x^{2}\Big[1-\frac{4}{3}\frac{1}{1+x^2}\Big]. 
\end{equation}
and  
\begin{equation}
\label{3.12}
\frac{|V'(x)|}{V}  = \frac{ \mid\frac{-2}{x} +12x+6x^3 \mid}{(-1+3x^2)(1+x^2)}=\frac{ \mid-2 +12c_S+6c_S^2 \mid}{\sqrt{c_S}(-1+3c_S)(1+c_S)},
\end{equation}
where Eq. (\ref{2.32b}) with $ \phi_0=M_P $ is used to get the second equality. Using the same argument used in the Cuscuton-like model, in the range  $ c_S \gg \mathcal{O}(1) $, Eq. (\ref{3.12}) reduces to 
\begin{equation}
\label{3.13}
\frac{|V'(x)|}{V}  \sim \frac{2}{\sqrt{c_S}}=2e^{-2N},
\end{equation}
which agrees Eq. (\ref{3.18}) with $ \beta=2 $ and shows that the second swampland conjecture is violated when  $ c_S \gg \mathcal{O}(1) $. Unlike the Cuscuton-like model, the DBI-like model is inconsistent with both de Sitter swampland conjectures. A plot showing Eq. (\ref{3.12}) and Eq. (\ref{3.13}) is shown in Fig. \ref{fig:2.0}.  

\subsection{C. The choice of $ \tilde{s} $ and field redefinition}

Since the background evolutions of the Hubble parameter and the speed of sound given in Eq. (\ref{2.16}) are independent of the choice of $ \tilde{s} $, an important question is whether any two different models can be related by a field redefinition $ \phi = f(\varphi) $. Here we show that for the master equation (\ref{2.10}) this is indeed this case, but we find that the Lagrangians which admit the given background evolution Eq. (\ref{2.16}) are not in general physically equivalent away from the attractor trajectory.  
 
First, by substituting Eqs. (\ref{2.17},\ref{2.19}) into  Eq. (\ref{2.10}), we obtain the special solution admitting the given background evolution as a special trajectory in phase space
\begin{equation}
\label{MasterEqofTC}
\dot{\phi}=\frac{\tilde{s}^2}{2}H(\phi)\phi, 
\end{equation}  
which also can be rewritten in the form of kinetic term as
\begin{equation}
\label{MasterEqofTC2}
X=\frac{\tilde{s}^2}{8}H^{2}(\phi)\phi ^2.
\end{equation}  
Eq. (\ref{MasterEqofTC2}) reduces to 
\begin{equation}
\label{CuscutonX}
X=\frac{s^2}{2}H^{2}(\phi)\phi^{2}
\end{equation}
for the Cuscuton-like model $ \tilde{s}=-2s $, and 
\begin{equation}
\label{DBIY}
Y=\frac{s^2}{8}H^{2}(\varphi)\varphi^{2}
\end{equation}
for the DBI-like model  $ \tilde{s}=s $, in which we use different notations $ \varphi $ for the field and $ Y=\frac{1}{2}\dot{\varphi}^2 $ for the kinetic term. The field redefinition $ \phi=f(\varphi) $ transforming Eq. (\ref{CuscutonX}) to Eq. (\ref{DBIY}) is given by the speed of sound 
\begin{equation}
\label{FieldRD}
c_S=\frac{\phi_0}{\phi}=\bigg(\frac{\varphi}{\varphi_0}\bigg)^{2},
\end{equation}
which should be invariant under field redefinition.
By substituting $ \phi=\frac{\phi_0\varphi_0^2}{\varphi^2} $ into Eq. (\ref{CuscutonX}), it is easy to show that 
\begin{equation}
\label{Fieldtrans}
X=\frac{1}{2}\dot{\phi}^2=\frac{1}{2}[\frac{d}{dt}\frac{\phi_0\varphi_0^2}{\varphi^2}]^2=\frac{s^2}{2}H^{2}(\frac{\phi_0\varphi_0^2}{\varphi^2})^{2}
\end{equation}
reduces to Eq. (\ref{DBIY}) by noticing that the Hubble parameter is another invariant quantity under field redefinition. 

However, it is important to realize that the field redefinition $ \phi=\frac{\phi_0\varphi_0^2}{\varphi^2} $ only holds on the special trajectory Eq. (\ref{MasterEqofTC}), but not for the Lagrangians $ \mathcal{L}(X, \phi) $(\ref{2.28}, \ref{2.29c}) since $ X $ and $ \phi $ are two independent variables for a general Lagrangian. An example to show that the two Lagrangians are physically distinguishable is the study of non-Gaussianity of the two models in \cite{Bessada:2012rf}, in which the two models having different signatures of non-Gaussianity was shown. Furthermore, it was shown in Ref. \cite{Bessada:2012zx} that the solution is a dynamical attractor in both the DBI and Cuscuton cases. 

The fact that two physically inequivalent Lagrangians can be chosen which exhibit {\it identical} background behavior demonstrates an interesting consequence of the freedom to choose more general kinetic terms in the Lagrangian: a field theory which violates {\it e.g.} the distance conjecture (\ref{1.1}) can be transformed to one which satisfies the conjecture via a field redefinition (or, equivalently, the choice of $\tilde{s}$) without altering the attractor solution for the background evolution. This calls into question whether or not the swampland conjectures (\ref{1.1},\ref{1.2}) should be applied unmodified to non-canonical theories, and indeed it is reasonable to expect that the conjectures would need to be generalized to cover such cases. Such a generalization was studied in the case of DBI inflation in Ref. \cite{Seo:2018abc}, in the ultra-relativistic limit $c_S \ll 1$, but would not be expected to apply the limit of $c_S > 1$. Given that there are no known string completions which exhibit $c_S > 1$, it is perhaps reasonable to add this condition itself as a swampland criterion, {\it i.e.} that UV-complete theories must have $c_S < 1$. (See Ref. \cite{Kinney:2007ag} for a discussion of this question.) Our goal in this work has been more modest, which is to check whether or not the unmodified conditions (\ref{1.1}) and (\ref{1.2}) place significant constraint on the consistency of cosmological models with $c_S > 1$, and we find that they do not. 

\section{V. conclusions}

In this paper, we consider tachyacoustic cosmology, an alternative to inflation theory with the speed of sound greater than the speed of light, in light of the widely discussed tension between the single-field inflation models and the de Sitter swampland conjectures \cite{Agrawal:2018own}. We have studied the two models introduced in Ref. \cite{Bessada:2009ns}, and have shown that the Cuscuton-like model is consistent with the swampland criteria (\ref{1.1}, \ref{1.2}) but the DBI-like model fails both conditions. We have further proposed a general scaling rule for the field excursion 
\begin{equation}
\label{key1}
\frac{\Delta \phi}{M_P} \sim \pm \sqrt{2\epsilon} (c_S)^{1/\beta} \Delta N, 
\end{equation}
where $\beta \equiv 2s/\tilde{s} $, which shows that $ \beta <0 $ is the condition to pass the first swampland criterion (\ref{1.1}) with $ c_S \gg 1 $. Next, under the condition that the potential has leading-order behavior of a power law in $ c_S $, we have shown that the logarithmic gradient of the scalar field potential scales as    
\begin{equation}
\label{key2}
M_P\frac{|V'(\phi)|}{V}\sim c_S^{-1/\beta}         
\end{equation} 
in the scale invariant limit with a radiation-dominated background, which shows that  $ \beta <0 $ is also the condition to pass the second swampland criterion (\ref{1.2}). Therefore, if the potential of the scalar field satisfies this general form of potential, the two swampland criteria set the same condition on tachyacoustic cosmology.  

We conclude that, at least in principle, models which solve the cosmological horizon problem via a superluminal sound speed $c_S \gg 1$, instead of accelerated expansion, can be consistent with the proposed de Sitter swampland conjectures and can therefore be extended to a UV-complete theory. It is not, however, clear that such models can be consistently embedded in string theory: for example, braneworld realizations of Dirac-Born-Infeld Lagrangians generically require $c_S < 1$ \cite{Kinney:2007ag}. Some efforts at embedding Cuscuton-like models have been made \cite{Compere:2011dx,Afshordi:2016guo}, but it is conceivable that tachyacoustic models require a context other than string theory for completion in the UV.\footnote{It is also not obvious whether models with $c_S > 1$ from thermal initial conditions \cite{Magueijo:2008pm,Afshordi:2016guo} have consistent UV completion, a question which we do not consider further in this work.} On the other hand, if such models do not have any consistent UV completion, it is the result of conditions other than proposed swampland conjectures.

\section*{Acknowledgments}
\begin{acknowledgments}
WHK is supported by the Vetenskapsr\r{a}det (Swedish Research Council) through contract No. 638-2013-8993 and the Oskar Klein Centre for Cosmoparticle Physics, and by the U.S. National Science Foundation under grant NSF-PHY-1719690. The authors thank Ghazal Geshnizjani for helpful conversation. 
\end{acknowledgments}

\bibliographystyle{apsrev4-1}
\bibliography{paper}

\end{document}